# Wettability of graphite under 2D confinement

*Zixuan Wei[1], Mara Chiricotto[1], Joshua D. Elliott[1], Fausto Martelli,[2,3] Paola Carbone[1*]*

[1]*Department of Chemical Engineering, The University of Manchester, Oxford Road M13 9PL, Manchester (UK)*

[2] *IBM Research Europe, WA4 4AD Daresbury, United Kingdom*
[3] *CNR-Institute of Complex Systems, Department of Physics, Sapienza University of Rome, P.le Aldo Moro, I-00185 Roma, Italy*

The thermodynamics of solid/liquid interfaces under nanoconfinement has tremendous implications for liquid transport properties. Here using molecular dynamics, we investigate graphite nanoslits and study how the water/graphite interfacial tension changes with the degree of confinement. We found that, for nanochannel heights between 0.7nm and 2.6nm, graphite becomes more hydrophobic than in bulk, and that the value of the surface tension oscillates before eventually converging towards a constant value for larger slits. The value of the surface tension is correlated with the slip length of the fluid and explained in terms of the effective and interfacial density, hydration pressure and friction coefficient. The study clearly indicates that there is a critical channel height of 0.9nm (achievable experimentally[1]) at which the surface tension reaches its highest value, but the water diffusion across the channel is at its minimum. The structural analysis shows that for this pore size a transition between a 2D and 3D hydrogen bond network is accompanied by an abrupt increase in conformational entropy. Our results show that the wettability of solid surfaces can change under nanoconfinement and the data can be used to interpret the experimental permeability data.

*Corresponding author. Tel: +441615293061 E-mail: paola.carbone@manchester.ac.uk (Paola Carbone)

1. Introduction

Understanding the physics of water under nanoconfinement underpins the development of many technologies in areas such as water desalination[2,3], energy storage[4,5] and heterogeneous catalysis[6]. The reason is that physical chemical properties of confined water are shown to be remarkably different than in bulk. Toney et al.[7] were one of the first groups that demonstrated, using X-ray scattering, that the density of liquid water in the immediate contact with a solid surface is higher compared to the bulk water. The authors also observed that the water molecules are ordered in layers extending about three molecular diameters from the surface. By changing the surface charge, they were able to identify the orientation of the molecules in direct contact with the surface. This molecular ordering at the solid/liquid interface becomes dominant under confinement. As confinement increases, the restrictions in the translational and rotational degrees of freedom that water molecules experience and the increasing dominance of the surface effects lead to specific molecular packing and reduce the local dynamics, driving dramatic changes in the fluid properties including (i) a decrease in dielectric constant of more than one order of magnitude[8], (ii) unexpected liquid-solid phase transitions[9], and (iii) confinement-specific water structures[10-13].

The structure of water under confinement has important consequences also on its flow properties and a significant effort has been invested in explaining the enhanced flux that water and electrolytes seem to achieve when pushed through nanopores. In general, the flux data are described by the Hagen-Poiseuille equation which describes the relationship between pressure, fluidic resistance and flow rate for Newtonian fluids subjected to laminar flow. The equation is normally employed assuming no-slip boundary conditions and bulk viscosity for water[14], however divergences between the theoretical predictions and experimental data are remarkable especially for nanopores. For instance Majumder et al.[15] experimentally measured the average velocity of water flowing through multiwalled carbon nanotubes with diameter of 7nm and found this velocity to be $10^4$ times greater than that predicted by conventional fluid flow theory. The following year, Holt et al.[16] reported enhanced flow of water through double-walled nanotubes of 1 – 2nm diameter of more than three orders of magnitudes while Whitby et al.[17] studied the water flux within 44nm carbon nanopipes and found this to be one order magnitude larger than that of the theoretical prediction. Recently, Radha et al.[18] have been able to assemble graphene 2D capillaries with sizes of angstroms precision. Also, in this geometry the authors observed fast water transport up to 1 m/s and attributed it to high capillary pressures and large slip lengths.

Various modifications of the basic Hagen-Poiseuille equation have been proposed over the years to explain the experimental data, including the use of an effective slip length[14] and the introduction of a material-dependent water/solid frictional interaction[19]. The need to include confinement-dependent values for various parameters within the equation is apparent as molecular simulations and experiments have indicated that almost all interfacial properties including viscosity[20], friction coefficients[21], interfacial slippage[22] and interfacial tension[23], are affected by both the size and geometry of the confinement.

While the structure of water in confinement has been extensively studied, how interfacial thermodynamic properties changes under different degrees of confinement and how this change may affect the water flux remains an open issue. Young et al.[24] experimentally found that the viscous shear forces of nanoconfined water dramatically changed between hydrophobic and hydrophilic surfaces. Giovambattista et al.[25] investigated how the structure of water nanoconfined between hydrophobic and hydrophilic plates change upon changing the capillary pressure. Wu et al.[26] employed Monte Carlo simulations to measure the local capillary pressure in the graphene channel as a function of confinement to relate its effect on water diffusion. More recently Calero and Franzese[27,24] investigated the relationship between the dynamical properties of and hydration pressure within single layer graphene 2D slits and showed that these are linked to the number of water layers within the pore.

In this work, we investigate whether the wettability of a graphitic surface changes under confinement and its relationship with the liquid/surface friction coefficients, the slip lengths and hydrogen bond network structure and entropy. The wettability is measured by calculating the solid/liquid surface tension which can be obtained integrating over the values of the local stress obtained from Molecular Dynamics (MD) simulations[28]. This procedure is normally carried out to study pressure gradient across soft liquid/liquid or liquid/gas interfaces such as lipid bilayers or micelles[29] but can also be used to measure the pressure inside solid confinements[26,27] and to estimate solid/liquid surface tension and work of adhesion[28,30-31]. To avoid forcing any specific density into the pore, we build our molecular models allowing the liquid water to fill graphitic 2D nanoslits of different heights and then we calculate the liquid/solid surface tension[30], friction coefficient[32] and slip length[33].

## 2. Methods and Systems

*2.1 Simulation details*

The 2D graphitic nanochannels are made of eight 10nm x 30nm rigid graphene sheets stacked on top of each other. A side-view snapshot of our system is shown in Figure 1. All the systems are periodic in all directions and thirteen different channel heights, $h$, between 0.7nm to 4.0nm are simulated. For each channel height, we build three uncorrelated initial configurations to test the reliability of the interfacial properties and calculate the standard deviations. The full list of studied systems is reported in Table 1.

Water is modelled using the TIP4P/2005 model[34]. This water model is known to accurately reproduce the bulk liquid/vapour surface tension[35,36] and has been used to study nanoconfined water[37,38,39] including to reproduce the water diffusion coefficient obtained from ab-initio MD[1]. The carbon atoms of graphene sheets are treated as neutral particles interacting with the oxygen atoms of the water through the LJ parameters $\sigma_c = 0.33997$nm and $\varepsilon_c = 0.3594$ kJ mol$^{-1}$ taken from Cornell et al.[40] The cross parameter interactions are computed using the Lorentz-Berthelot combining rules: $\delta_{ij} = 0.32793$nm and $\varepsilon_{ij} = 0.52804$ kJ mol$^{-1}$. These LJ parameters reproduce the experimental water-graphene surface tension[30,41,42] and predict a contact angle in agreement with recent experiments [41–43]. To generate our systems, we create a slit channel of a specific height, $h$, of pristine 2D flat graphitic surfaces ($10 \times 30$nm) in contact with two reservoirs of bulk water of around 5nm thickness, then we run a 20ns NVT molecular dynamics simulation during which the water molecules from the reservoirs are allowed to diffuse and fill the slit. During the NVT simulations the water reservoirs are in equilibrium with a vapour phase (Figure 1). The liquid/vapour interface is needed to ensure that the density of the water reservoirs is that of bulk water. This procedure to fill the slit ensures that the density of water in the pore is the correct for these thermodynamic conditions[1]. At this point, the reservoirs are removed, and periodic boundary conditions are applied to the edges of graphene sheets, as shown in Figure 1, this system is further equilibrated for 60ns[37]. We hold T constant at 298K with the Nosé/Hoover thermostat[46] with relaxation times of 0.5ps. The equation of motion is integrated with the Verlet algorithm with a time step of 0.002ps. The Lennard-Jones potential smoothly taken to zero between 1.0 nm and 1.2 nm using a spline switching function. The long-range electrostatic forces are computed using the particle mesh Ewald (PME) with a grid space of 0.12nm and a Fourier grid spacing of 0.15nm. We use the Gromacs (v 2018.4) software package[47] to carry out all the molecular dynamics simulations.

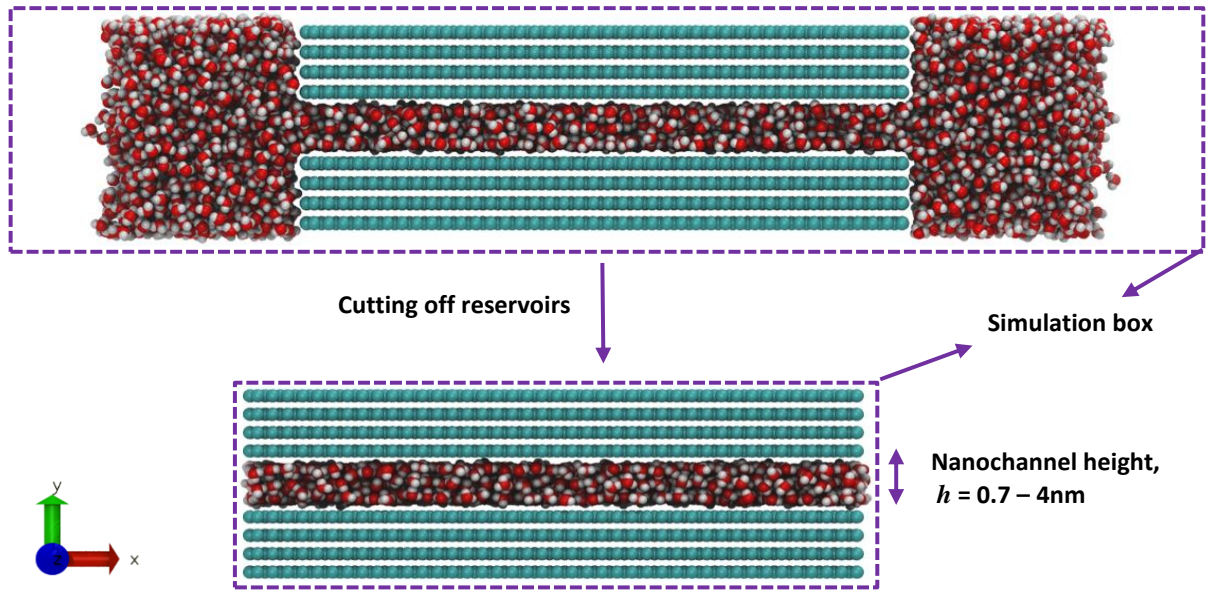

Figure 1. Side-view of water confined in a graphitic nanochannel of 1.2nm height: snapshot of the channel in contact with the two water reservoirs after the water has filled the channel (top); snapshot of the channel after the reservoirs have been removed (bottom). The y-direction is normal to the graphite surface. Periodic boundary conditions are applied in all three directions. The surface area is defined by A = LxLz. The red, white, cyan spheres represent oxygen, hydrogen, and carbon atoms respectively. In the top panel the size of the vapour phase (which is in length 3 times the thickness of the water reservoir) has been reduced for clarity.

*2.2 Calculation of the interfacial tension from the local stress profile*

The solid/liquid interfacial tension is calculated using its mechanical definition, where the surface tension is obtained from the local stress tensor. Using the values of the atomic forces and velocities calculated during the MD simulation, the local stress tensor, $\sigma(x)$, can be obtained based on the Irving-Kirkwood formalism[48] which defines $\sigma(x)$ as the sum of the kinetic, $\sigma_K(x)$, and intermolecular potential, $\sigma_V(x)$, components as:

$$\sigma(x) = \sigma_K(x) + \sigma_V(x) \qquad (1)$$

Where $x$ is the centre of a three-dimensional rectangular element of a grid used to discretize the simulation box. The expression of $\sigma_K(x)$ and $\sigma_V(x)$ are

$$\sigma_K(x) = - \left\langle \sum_{\alpha=1}^{N} m^\alpha v^\alpha \otimes v^\alpha \delta(r^\alpha - x) \right\rangle \quad (2)$$

$$\sigma_V(x) = \left\langle \sum_{\alpha,\beta>\alpha}^{N} f^{\alpha\beta} \otimes r^{\alpha\beta} B(r^\alpha, r^\beta; x) \right\rangle \quad (3)$$

where $r^{\alpha\beta}$ and $f^{\alpha\beta}$ are the intermolecular position vector and intermolecular force vector between particle $\alpha$ and $\beta$ respectively, $m$ and $v$ are the mass of the particle and its velocity vector. For a system with planar symmetry and isotropic surface, such as in the present case, the tensor $\sigma(x)$ can be reduced to its three components: $\sigma_{xx}$ and $\sigma_{zz}$, parallel to the surface in this set up, and $\sigma_{yy}$ which is the component of the stress perpendicular to the surface. Thus we define the normal component of the pressure tensor along this direction as $P_N(y) = -\sigma_{yy}(y)$, and $P_T(y)$, which corresponds to the tangential pressure, is given by $P_T(y) = -\frac{1}{2}(\sigma_{xx}(y) + \sigma_{zz}(y))$. Hence, we can compute the interfacial tension $\gamma$ through the calculation of integral of the stress profile[28]:

$$\gamma = \int_{-h/2}^{h/2} (P_N(y) - P_T(y)) \, dy \quad (4)$$

where $h$ is the channel height. The method is widely employed to compute gas/fluid and fluid/fluid surface tensions[35,49], but recently it has been used also for solid/liquid interfaces[30]. The details of the method can be found in several publications[41,50,51,52] including our recent work where we followed the same procedure to calculate the work of adhesion between polyisoprene and graphite[28] and the graphene/water interfacial tension[30]. The GROMACS-LS[53] (version 2016.3) was used to compute the local stress profile following the same procedure as detailed in Ref. [30].

2.3 Calculation of the slip length

To evaluate the slip length we follow two different procedures using both Equilibrium Molecular Dynamics simulation (EMD) and the Non-Equilibrium Molecular Dynamics simulation (NEMD)[54].

From the EMD simulations, we compute the slip length $L_s$ by using the Navier relation[55]:

$$L_s = \frac{\eta_0}{\lambda} \quad (5)$$

where $\lambda$ is the solid-liquid interfacial friction coefficient calculated via EMD simulations as described below and $\eta_0$ is the water bulk shear viscosity, which for the TIP4P/2005 is $\eta_0$=0.855 mPa·s at 298K[56].

To assess the robustness of the results, for some of the larger channels using NEMD simulations by applying a constant acceleration (in-plane direction) on each atom of the water molecules in the nanochannels[33] aiming to generate a Poiseuille flow, we then calculate $L_s$. Due to the limitation of the length of the simulations, on the order of tens of nanoseconds, a high acceleration must be applied to the water atoms to generate a high enough streaming velocity to calculate the slip length. For our specific set up we use an acceleration ranging from 2×10$^{11}$ m/s$^2$ to 2×10$^{12}$ m/s$^2$ applied to all oxygen and hydrogen atoms. This creates a pressure gradient between 2×10$^{14}$ Pa/m to 2×10$^{15}$ Pa/m, which represents a good compromise between computational efficiency and being within the linear regime limit[57]. With these acceleration values, a slip velocity of around 14m/s is obtained. Similar pressure gradient values and corresponding slip velocities were reported by Kannam et al.[33, 49] and Kotsalis et al.[59]. To calculate the slip length, the flow velocity profiles, $u_x(y)$, are then fit to a quadratic equation [58,60,61]:

$$u_x(y) = ay^2 + b \qquad (6)$$

where $b$ is the fitting parameter, the value of $a$ is constrained to the one that reproduces the bulk viscosity of the fluid, $\eta_0$, and $u_x$ are the $x$-component of the oxygen atoms' velocities. This constrained fitting procedure has shown to be able to reproduce a slip length comparable to that obtained from EMD[33]. The slip length is then calculated using:

$$u_s = L_s \left.\frac{\partial u_x}{\partial y}\right|_{y=y_{wall}} \qquad (7)$$

where $u_s$ and $L_s$ are the slip velocity and the slip length respectively and $\left.\frac{\partial u_x}{\partial y}\right|_{y=y_{wall}}$ is the strain rate at the wall ($y_{wall}$). Examples of average streaming velocity profile of the oxygen atoms along $y$ axis averaged over 20 independent NEMD simulations are reported in the supporting information (Figure S1), each NEMD simulation is 20ns with time step of 0.002ps under NVT ensemble.

## 2.4 Calculation of the Friction coefficient

The water/graphite friction coefficients are also calculated from EMD simulations using the Green-Kubo relationship[32]

$$\lambda = \frac{1}{A\, k_B T} \int_0^t \langle F(t)F(0)\rangle dt \qquad (8)$$

Where $\lambda$ is the friction coefficient, $F(t)$ is the total tangential (i.e. in the $xz$ plane in Figure 1) force acting between the graphitic surface and the water at time t, A is the total surface area of the slit pore and the average runs over the length of the trajectories. The integral of the autocorrelation function is calculated at *t*=10ps for all systems and the forces are saved every 0.2ps. Five independent 3ns long EMD simulations are used to calculate $\lambda$ and its standard deviation. Due to the computationally demanding simulations and large output files produced, the majority of the simulations to calculate $\lambda$ are performed on a reduced graphitic surface of 5nm × 10nm. We verified that the value of the calculated $\lambda$ is independent on the graphitic sheet size.

## 2.5 Calculation of the 2D self-diffusion coefficient

The lateral self-diffusion coefficient, $D_\parallel$, is calculated from the mean square displacement (MSD) via the Einstein relation:

$$D_\parallel = \lim_{t\to\infty} \frac{1}{4Nt} \langle \sum_{i=1}^{N}[r_i(t) - r_i(0)]^2 \rangle \qquad (9)$$

where $r_i(0)$ and $r_i(t)$ are the positions of the molecules at time 0 and $t$, respectively and $N$ is total number of water molecules. The vector $r_i$ contains only the in-plane components of the $i$th atomic position and the self-diffusion coefficient $D_\parallel$ is calculated only in the plane parallel to the surface.

## 2.6 Hydrogen bonds structural analysis

In network-forming materials like water, the arrangement of the network of bonds contribute to the property of the system in terms of configurational entropy[62]. Therefore, assessing the topology of the network of bonds is a valuable tool to describe the properties of water. Broadly speaking, the network topology in disordered materials is an important structural descriptor for understanding the nature of the disorder that is usually hidden in pairwise correlations. Correlations between dynamics, anomalous behaviour and the Hydrogen Bond

Network (HBN) have been reported in the case of bulk water[63,64] and in water under soft and hard confinement[30,65,66]. The study of the HBN topology can be performed via the ring statistics, which inspect the HBN looking for closed loops (or rings) formed by hydrogen bonds. In order to compute the ring statistics, it is necessary to define the link between atoms/molecules. Possible definitions can be based on the formation of bonds, interaction energies, geometric distances, etc.. The second step is the definition of a ring and the corresponding counting scheme. This task is of relevance in directional networks, like water or silica, were the donor/acceptor nature of the bonds breaks the symmetry in the linker search path. In this work, we adopt the following counting scheme which has been shown to carry the information about the HBN[66] and to directly connect its topology with properties of water such as translational diffusion, rotational dynamics and structural properties[63]: starting from a water molecule we construct rings recursively traversing the HBN until the starting point is reached again or the path exceeds the maximal ring size considered (12 water molecules in our case). We consider primitive rings only, i.e., rings that cannot be reduced to smaller ones. In this study, we do not discern among the acceptor/donor character of the starting water molecule. The distribution has been computed in the region of the first peak in density.

## 3. Results and Discussion

*3.1 Surface tension and channel density*

We initially calculate the interfacial density, $\rho_{\text{interfacial}}$, and effective density, $\rho_{\text{effective}}$, within the channels (Table 1, column two and three). The effective density is defined as:

$$\rho_{\text{effective}} = \frac{n \, M_{\text{water}}}{N_A (A \cdot h_{\text{effective}})} \quad (10)$$

where $n$ is total number of water molecules in the channel, $N_A$ is the Avogadro constant, $A$ is the graphene surface area, $h_{\text{effective}}$ is the effective height of channel, which is the channel height ($h$) minus $\sigma_c$, and $M_{water}$ is the molar mass of the water molecule. The interfacial density, $\rho_{\text{interfacial}}$, is instead the water density near the graphitic surface defined as the value of the first density peak in the cross-sectional density profiles.

*Table 1. Effective and interfacial density and surface energy for slit pores of different heights, h.*

| $h$ (nm) | $\rho_{effective}$ (kg/m³) | $\rho_{interfacial}$ (kg/m³) | $\gamma$ (mN/m) |
|---|---|---|---|
| 0.7 | 908.4 | 3801.3 | 14.3 ± 0.1 |
| 0.8 | 858.8 | 2559.7 | 39.7 ± 1.6 |
| 0.9 | 1060.2 | 6029.2 | 429.3 ± 0.3 |
| 1.0 | 960.5 | 3056.0 | -58.4 ± 0.2 |
| 1.2 | 989.3 | 3708.7 | 122.9 ± 0.5 |
| 1.6 | 984.4 | 3462.3 | 71.2 ± 0.5 |
| 1.8 | 978.5 | 3293.8 | 107.1 ± 0.6 |
| 2.0 | 972.1 | 3296.1 | 138.0 ± 0.5 |
| 2.3 | 978.9 | 3319.0 | 103.9 ± 0.7 |
| 2.6 | 978.9 | 3255.9 | 69.5 ± 0.5 |
| 3.0 | 975.7 | 3316.9 | 66.2 ± 0.8 |
| 3.5 | 977.6 | 3335.0 | 68.3 ± 1.2 |
| 4.0 | 976.5 | 3350.3 | 67.6 ± 0.8 |

In Figure 2, we report the cross-sectional density profiles for channels with heights equal to 0.7, 0.9, 1.0, 2.0 and 4.0nm for which we observe the most representative water layering structures. The density profiles for the other channel heights are reported in Figure S2 in the supporting information. The simulations show that for $h$=0.7nm, the water molecules are arranged in a single layer placed in the middle of the channel. This monolayer starts to expand and splits in two for $h$=0.8nm. When the channel height reaches 0.9nm, two distinct water layers are formed, and the interfacial density reaches its highest value (see Table 1). As the channel height further increases the number of water layers increases as well and the interfacial density values plateau at 3300kg/m³ ca., while the density in the middle of the channel equal to the bulk. The oscillations observed in the effective density values are consistent with those reported in previous works[25,55–57] for similar thermodynamic conditions, although, due to the different surface/water force field parameters, the water model employed and the simulation set up, the interlayer distances at which they are observed and the actual density values differ. Table 1 reports also the graphite/water interfacial tension values calculated for all channel heights, the values are also plotted in Figure 3.

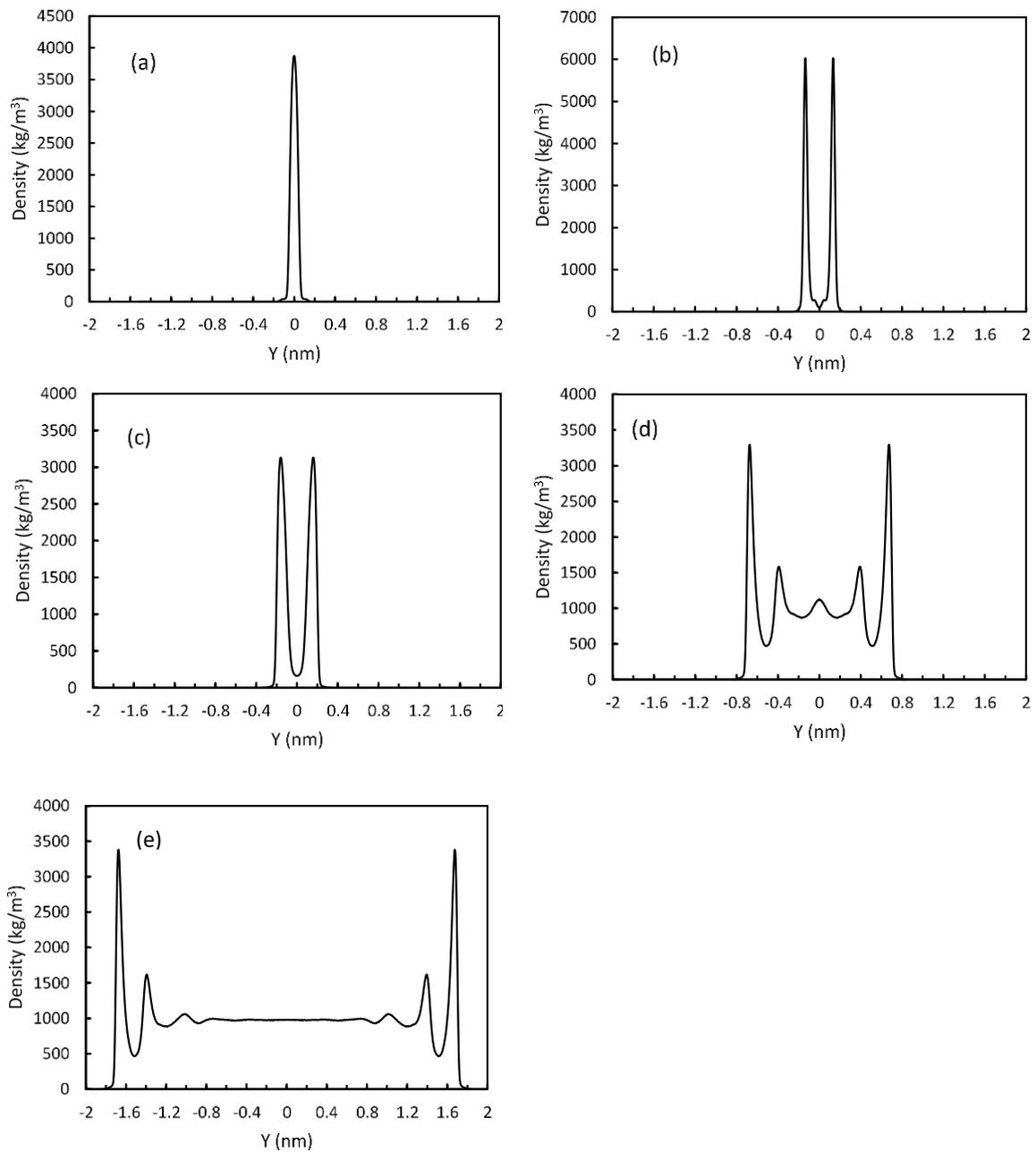

*Figure 2. Cross-sectional density profiles as a function of the Y (y-axis) which is the perpendicular direction for different channel heights 0.7nm (a), 0. 9nm (b), 1nm (c), 2.0nm (d) and 4.0nm (e), Y=0 corresponds to the centre of the channel.*

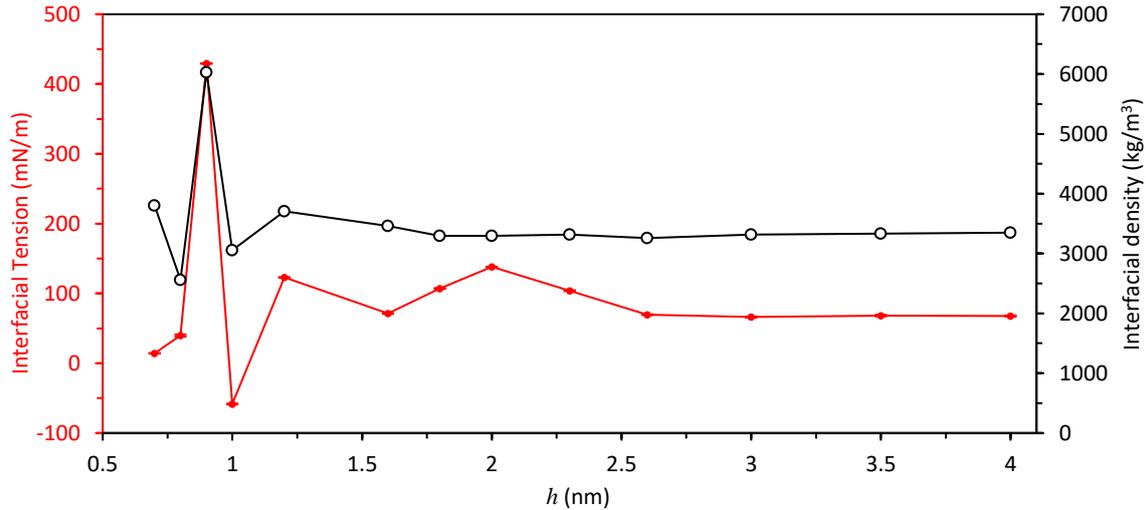

*Figure 3. Interfacial tensions with the standard deviation (red solid line with red dots) and the interfacial density (black solid line with black empty circles) as function of channel height, h. Lines are added to guide the eye.*

Inspection of Figure 3 reveals that, except for $h$=0.7nm, there is a correlation between the interfacial density and the surface tension at least for values of $h \leq 1.6$nm. The data indicate that as the interfacial density peaks at $h = 0.9$nm, the interfacial tension follows suit and that graphitic pores of this size are around 4 (10) times more hydrophobic than graphitic pores of larger (smaller) size. The negative interfacial tension observed for $h = 1.0$nm is an indication of the mechanical instability of channels of this height which, if allowed to relax, would collapse to smaller sizes[1]. The surface tension peaks again for $h = 2.0$nm, although the interfacial density value remains instead constant for all $h$ values larger than 1.6nm.

We can then classify the level of confinement in three regimes: strong confined channel ($h = 0.7$–1.0nm), intermediate confined channel ($h = 1.2$–2.6nm) and bulk-like channel ($h = 3.0$–4.0nm). Under strong confinement, the channel interfacial tension dramatically increases with the rapid increase of the interfacial density (except for channels of height 1.0nm, which are mechanically unstable). At intermediate confinement, a second peak in interfacial tension at $h = 2.0$nm occurs. This does not however correspond to a peak in interfacial density, but to the appearance of a fifth peak in the cross-sectional density profile with density still higher than the bulk water density (Figure 2d). For channels under bulk-like confinement, the interfacial tension plateaus around γ=67mN/m. This value is similar to but still 20% larger than the graphite/water interfacial tension calculated by Chiricotto et al. [30] in unconfined systems. This result indicates that despite at this confinement both interfacial and effective densities plateau, bulk conditions have not been reached yet as also shown by the fact that effective densities are still below the bulk density value.

The oscillatory behaviour shown by the surface tension as a function of level of confinement as reported in Figure 3, is related to the value of the hydration pressures ($P_{hyd}$) within the channel. To calculate $P_{hyd}$ we calculate the average normal component of the pressure $\overline{P_N}$ from the local stress profile as:

$$\overline{P_N} = \frac{\sum_{i=1}^{n} -\sigma_{yy}(i)}{n} \tag{11}$$

where the $\sigma_{yy}$ is the component of the stress tensor normal to the plane of the graphitic sheets (i.e. $yy$), $n$ is the number of bins along the whole simulation box, including both the water reservoirs and the channel (see Figure 1) in the $z$ direction. A bin size of 0.1nm is used in the calculation. The hydrostatic pressure is then calculated as the difference between the normal pressure calculated inside the channel, $\overline{P_N^C}$, and that calculated in the reservoir, $\overline{P_N^R}$. For large channels, we expect the two pressures to be almost the same and $P_{hyd} \approx 0$, as confinement increases the two pressure values will differ and, based on the surface tension results, we expect an oscillatory behaviour. It is worth noticing here that the pressure in the reservoir is not the atmospheric pressure as one might expect but $\overline{P_N^R}$ is around $-100$MPa due to the small size of the water reservoirs which extend in the $x$ direction for 3nm. Despite this, the values of the surface tension (calculated as the integral of the difference of the normal and tangential pressure) are not affected. As expected, the value of hydration pressure oscillates as $h$ changes, (Figure 4) and is negative (indicating channel collapsing) only for $h$=1.0nm for which $\gamma$ is negative. For the other channel heights investigated here $P_{hyd}$ is always positive with maxima for $h = 0.9$, 1.2 and 2nm as found for $\gamma$. This oscillatory behaviour is in agreement the results obtained by Striolo and co-workers[68] that calculated the solvation pressure in graphitic slit pores resembling the ones modelled here and by Calero and Franzese[27], although in the latter case the pore is a small graphene-slit immersed in bulk water. In both cases these oscillations, which occur at slightly different channel heights due to the different surface/water interactions, method to calculate the pressure and simulation set up, occur in parallel with changes in the water layering structure in the pores.

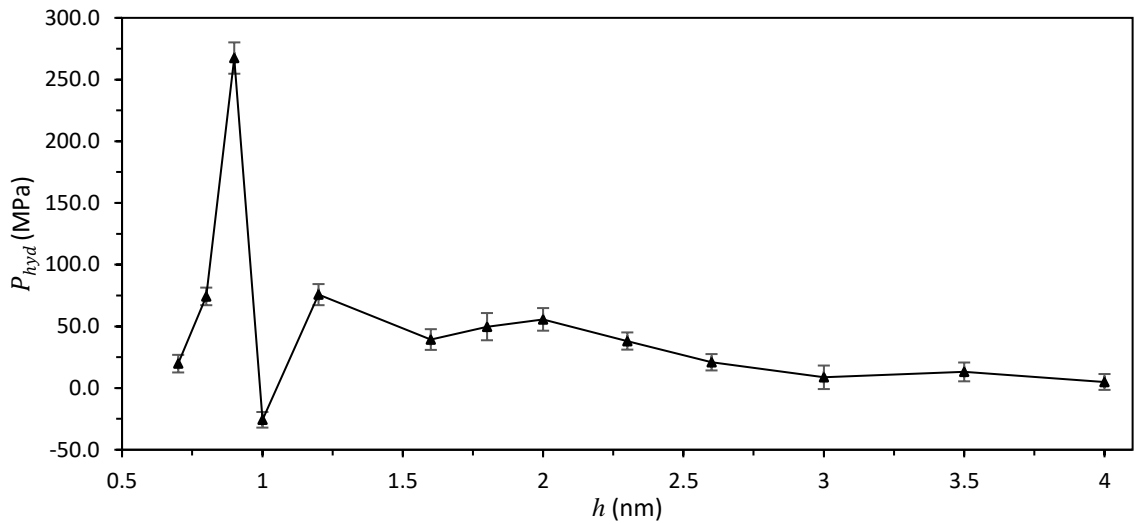

*Figure 4. Change in the hydration pressure as a function of channel heights, h.*

*3.2 Slip length and friction coefficient*

Table 2 reports the values of the friction coefficients, $\lambda$, and slip lengths, $L_s$, for each channel height. For some of the channel sizes, the values of $L_s$ calculated from $\lambda$ via equation 5 (second column of Table 2) are compared with those obtained from the velocity profiles obtained from the NEMD simulations. This comparison is feasible only for the largest channels where the problem of the uncertainty related to the fitting procedure of the flat velocity profiles normally obtained from the NEMD simulations[58,33,60] can be overcome by using the constrained fitting procedure described in the methodology section. For channel heights smaller than 3.0nm, the velocity profiles calculated across the channel are too flat to be fitted (see Figure S1 (a) in the Supporting Information) and therefore for these cases the only EMD results are reported. The agreement between the data obtained with the two procedures is good and demonstrates the robustness of the EMD approach followed to calculate the friction coefficients also in the strong confinement regime.

Table 2 *The EMD and the NEMD slip lengths and friction coefficients calculated for different channel heights, h.*

| $h$ (nm) | $L_S^{(EMD)}$ (nm) | $L_S^{(NEMD)}$ (nm) | $\lambda$ ($1\times10^4$ Nsm$^{-3}$) |
|---|---|---|---|
| 0.7 | 52.7 ± 3.3 | - | 1.68 ± 0.10 |
| 0.8 | 70.2 ± 2.4 | - | 1.26 ± 0.43 |
| 0.9 | 27.5 ± 3.2 | - | 3.22 ± 0.37 |
| 1.0 | 38.9 ± 3.9 | - | 2.20 ± 0.25 |
| 1.2 | 47.5 ± 7.1 | - | 1.86 ± 0.28 |
| 1.6 | 55.9 ± 8.8 | - | 1.58 ± 0.25 |
| 1.8 | 50.9 ± 4.1 | - | 1.74 ± 0.14 |
| 2.0 | 51.1 ± 5.2 | - | 1.73 ± 0.18 |
| 2.3 | 48.2 ± 4.6 | - | 1.84 ± 0.18 |
| 2.6 | 49.0 ± 7.0 | - | 1.81 ± 0.03 |
| 3.0 | 46.2 ± 7.0 | 56.9 ± 5.6 | 1.92 ± 0.29 |
| 3.5 | 50.7 ± 1.9 | 50.6 ± 7.5 | 1.75 ± 0.06 |
| 4.0 | 50.1 ± 1.1 | 41.5 ± 5.6 | 1.77 ± 0.40 |

Under strong confinement, large oscillations of the slip length and the friction coefficient occur. The smallest slip length is for $h$=0.9nm, which is the pore size with largest friction coefficient and surface tension; the largest value is calculated for $h$=0.8nm corresponding to the smallest friction coefficient and low surface tension, although not the lowest which instead occurs for $h$=0.7nm. Figure 5 shows that, under strong confinement, the friction coefficient follows the same trend with $h$ as the interfacial density.

As observed for the interfacial density, slip lengths and friction coefficients plateau for $h$>2.0nm, and the value they converged to agree with previous simulation data.[22,33,71] This behaviour is also mirrored by the structuring of the water within the channel.

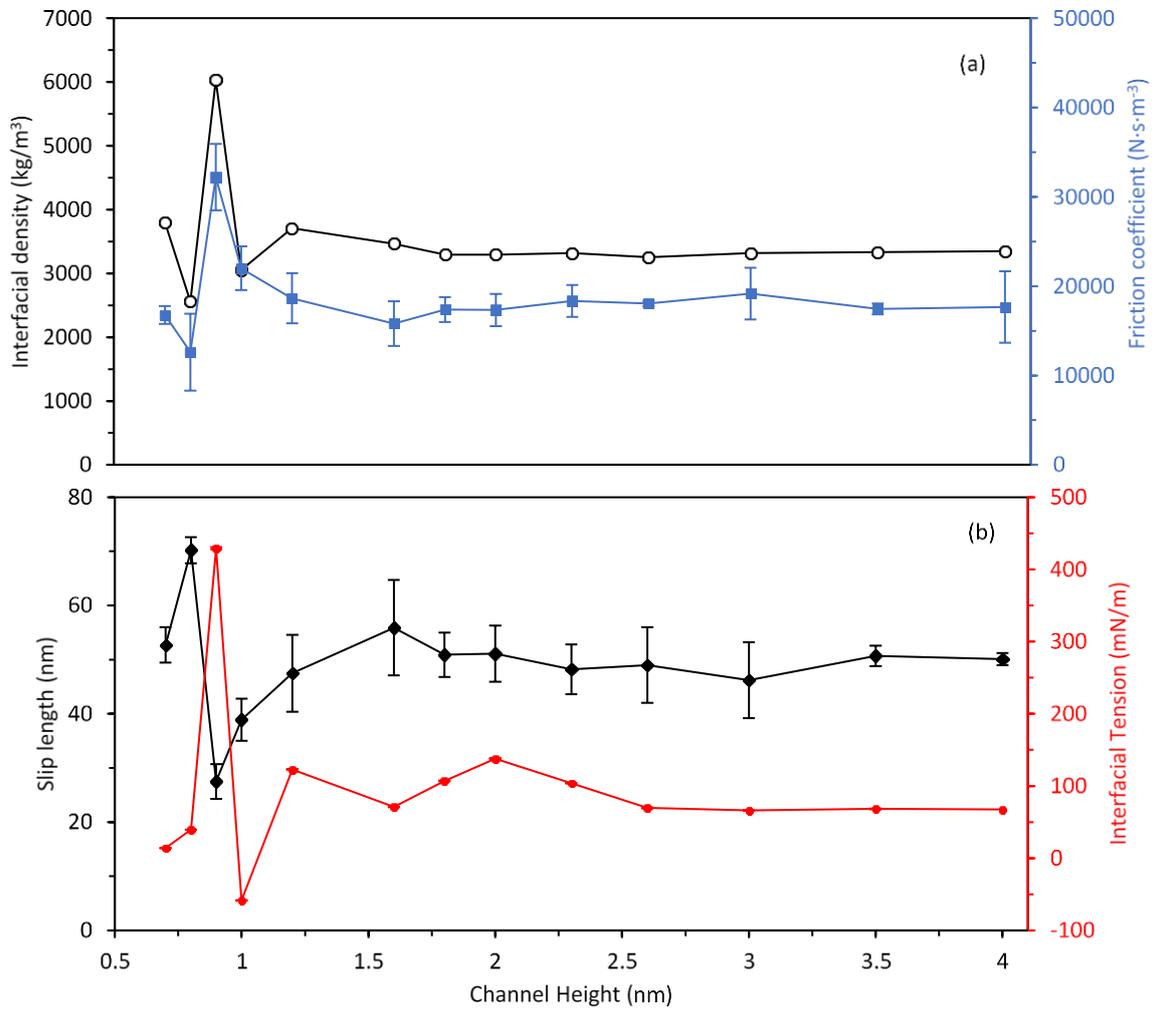

*Figure 5 (a). The friction coefficient (blue solid line with blue filled squares) with the standard deviation and the interfacial density (black solid line with black empty circles) against with the channel heights. (b) The slip length (black solid line with black filled diamonds) with the interfacial tension (red solid line with red dots) and associated standard deviation as a function of the channel heights.*

*3.3 Diffusion coefficient and HBN structural analysis*

The relation between slip length and surface tension is of fundamental importance to understand water flow in nanopores as both values enter into the Hagen-Poiseuille equation and its various modifications[14]. Here we want to highlight that our results are only in apparent disagreement with the relationship proposed by Huang et al[72] indicating that the slip length increases with the contact angle or in other words, increase with decreasing wettability. The argument used by Huang and co-authors to justify the relationship assumes that the diffusion coefficient of the fluid is constant (as it is in bulk). However, within the channel the in-plane lateral water self-diffusion coefficient (Figure 6) fluctuates as a function of the channel height following the same trend showed by the slip length (Figure 5b). The fastest diffusion is observed for the 0.7nm channel, while the slowest lateral diffusion coefficient is obtained for $h$=0.9nm and the second smallest value is obtained for the 1nm channel, result that agrees with the lateral diffusion coefficient calculated by Calero and Franzese[27]. The lateral self-diffusion coefficient increases with decrease of the effective density under the strong confined regime and plateaus to the bulk TIP4P2005 value ($2.4 \times 10^{-5}$ cm$^2$ s$^{-1}$) for for $h$≥1.2nm. All data of lateral self-diffusion coefficients are reported in Table 3.

*Table 3 The lateral self-diffusion coefficient calculated for different channel heights, h.*

| $h$ (nm) | $D_\parallel$ ($1\times10^{-5}$ cm$^2$/s) |
|---|---|
| 0.7 | 4.09 ± 0.22 |
| 0.8 | 2.63 ± 0.26 |
| 0.9 | 1.53 ± 0.02 |
| 1.0 | 1.78 ± 0.02 |
| 1.2 | 2.25 ± 0.17 |
| 1.6 | 2.33 ± 0.10 |
| 1.8 | 2.31 ± 0.06 |
| 2.0 | 2.37 ± 0.03 |
| 2.3 | 2.24 ± 0.08 |
| 2.6 | 2.33 ± 0.01 |
| 3.0 | 2.24 ± 0.04 |
| 3.5 | 2.37 ± 0.11 |
| 4.0 | 2.26 ± 0.03 |

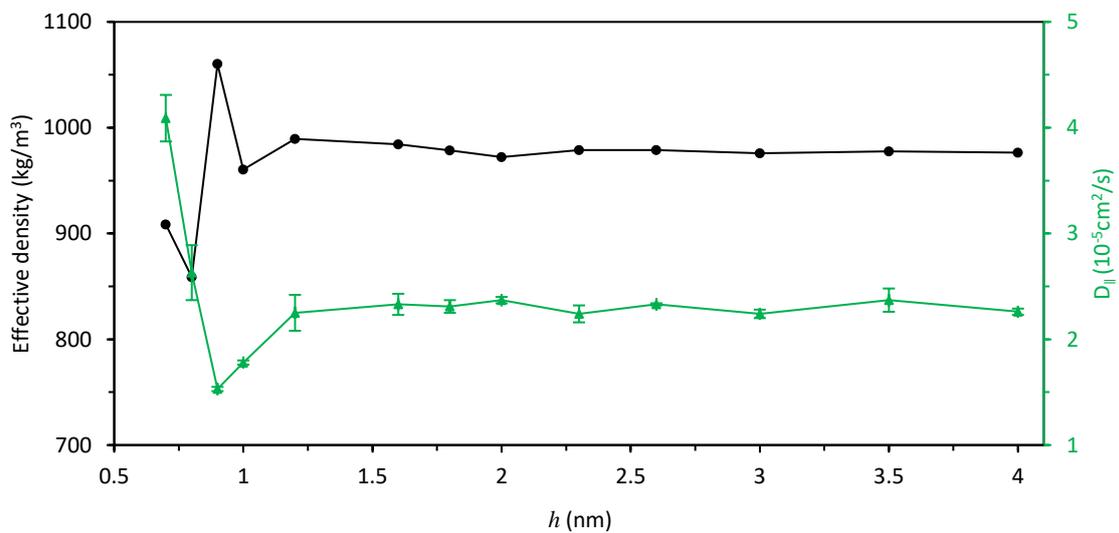

*Figure 6. The lateral self-diffusion coefficient (green solid line with green filled triangles) and the effective density (black solid line with black filled circles) against with the channel heights.*

In the strongly confined channels containing discrete water layers, the in-plane water structure at the interface is well characterized by the two-dimensional radial distribution function (2D RDF) between the oxygen atoms calculated as[37]

$$g_{II}(r_{xz}) = \frac{1}{2\pi\rho y r_{xz} \delta r} \sum_{n=0}^{b} \sum_{i \neq j} (H(n\delta r - r_{xz}) - H((n+1)\delta r - r_{xz}))$$

$$H(x) = \begin{cases} 1 & x > 0 \\ 0 & x \leq 0 \end{cases} \quad (12)$$

where the $\rho$ and $r_{xz}$ are the density of oxygen atoms and the distance of $i$ and $j$ oxygen atoms in the *xz*-plane, respectively, $b$ is the number of bins and $H$ is the Heaviside step function. In Figure 7 we report the oxygen-oxygen 2D-RDFs calculated for only the first interfacial water layer adjacent to the graphitic surface. The plots (Figure 7) show that the 2D structure of the first interfacial layer of water is different under the strong confinement ($h$=0.7–1.0nm) compared to larger channels for which the RDFs are all very similar (Figure S3 in the SI). In particular, the plots indicate that in the narrowest channels the water molecules are more packed (first peak of the RDF is sharper) with the exception of the $h$=0.8nm channel, which shows the smallest interfacial density. To understand better the structural transition occurring around $h$=1.0nm, we performed a HBN (Hydrogen Bond Network) analysis of the confined water.

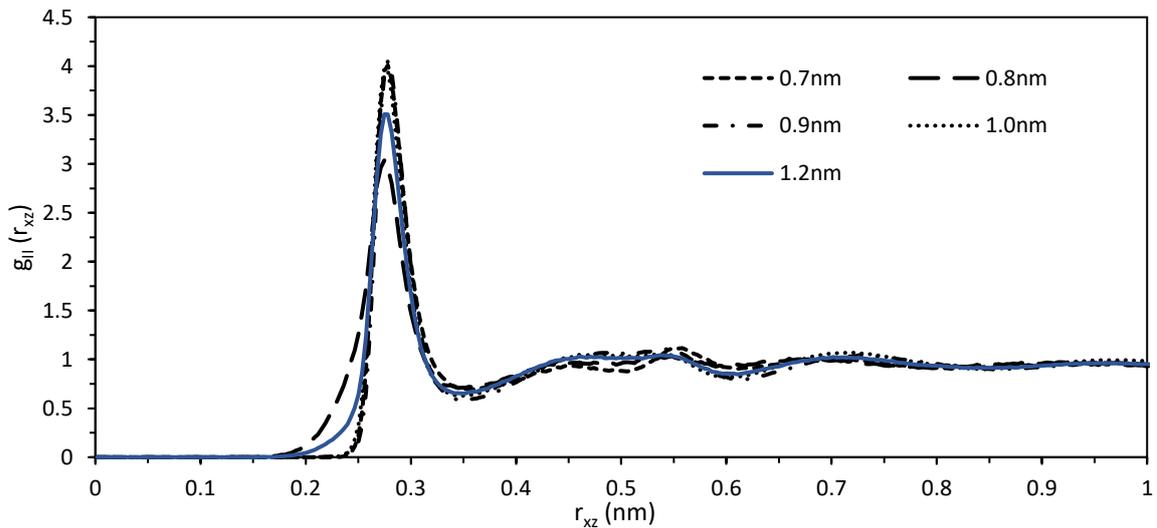

*Figure 7. The oxygen-oxygen 2D RDFs, $g_{II}(r)$, calculated for channel heights between 0.7 and 1.0nm and (for comparison) for $h$=1.2nm.*

In figure 8 (a) we report the distribution of HBN rings as a function of the degree of confinement. In the strongly confined regime ($h<1.0$nm) we can observe that the ring distribution strongly deviates from that typical of bulk water, in agreement with the behaviour of the interfacial tension. It is interesting to notice that while in bulk conditions longer rings (n>8) account for higher density[63,73], under confinement the high interfacial density is associated to very low number of longer rings and high number of shorter rings. In particular, we observe a strong contribution of square rings. Squared rings are stiff configurations that increase the spread of three body bond angles and are hard to anneal[74]. Moving towards lower confinements, we observe that the case $h=0.9$nm shows a bimodal distribution with a peak at n=4 and a peak at n=6. This bimodal distribution is remarkable, signalling a topological phase transition between a quasi-2D network characterizing strong confinements ($h<0.9$nm) to a full 3D network of lower confinements ($h>0.9$nm). This transition involves a massive rearrangement of the network, with the sudden increase, with respect to $h<0.9$nm, of n=6 and n=7 and a corresponding reduction of n=4. This transition occurs because the density peak corresponding to $h=0.9$nm is large enough to accommodate water molecules sitting in an intermediate configuration between a quasi-2D system and a 3D system. Thus, the lower lateral diffusivity observed for $h=0.9$nm is caused by the opening of the third dimension which allows water molecules to diffuse also perpendicularly to the surface. Calculating the mean-squared displacement (MSD) of the water molecules perpendicular to the graphite plane, Figure S4 in the SI, we notice that for $h<0.9$nm the water molecules motion is restricted in the $xz$ plane but as the channel height increases, water starts to move also in the third dimension. Upon further increasing of the distance between the plates, the main density peak spreads further and can accommodate more water molecules and transitions towards a full 3D network. For $h=1.0$nm the P(n) shows a reduction of n=4 and an increase in n=5, causing the disappearance of the bimodality observed for $h=0.9$nm. Upon further increasing the distance between the surfaces, we observe a convergence of the network topology towards values similar to those of bulk water[63], with an increase of longer rings (n>8) signalling the ability of water molecules to arrange in the third dimension.

Considering that the topology of the HBN affects the configurational entropy $S_c$[62], we estimate the contribution of $S_c$ as follows:

$$\frac{S_c}{k_B} = \sum_{i=3}^{12} p_i \ln p_i \qquad (13)$$

where the sum runs over all computed rings lengths and $p_i$ represents how many times a given ring occurs and $k_B$ is the Boltzmann constant.

In figure 8 (b) we report the profile of $S_c$ as a function of the confinement. It is worthy to remark, at this point, that Franzese et al. have shown that the transition from a 2D to a 3D system is driven by enthalpy in a system composed by two graphene flakes solvated in water[27]. With respect to Franzese et al[27], in this work we have direct access to the spatial arrangement of the HBN. As shown in figure 8 (b), in correspondence with the 2D-to-3D transition occurring at $h$=0.9nm, we observe a sudden jump in $S_c$, suggesting that this topological transition is entropic in nature and further proving that this level of confinement allows the network to explore a third dimension otherwise inaccessible at lower confinements.

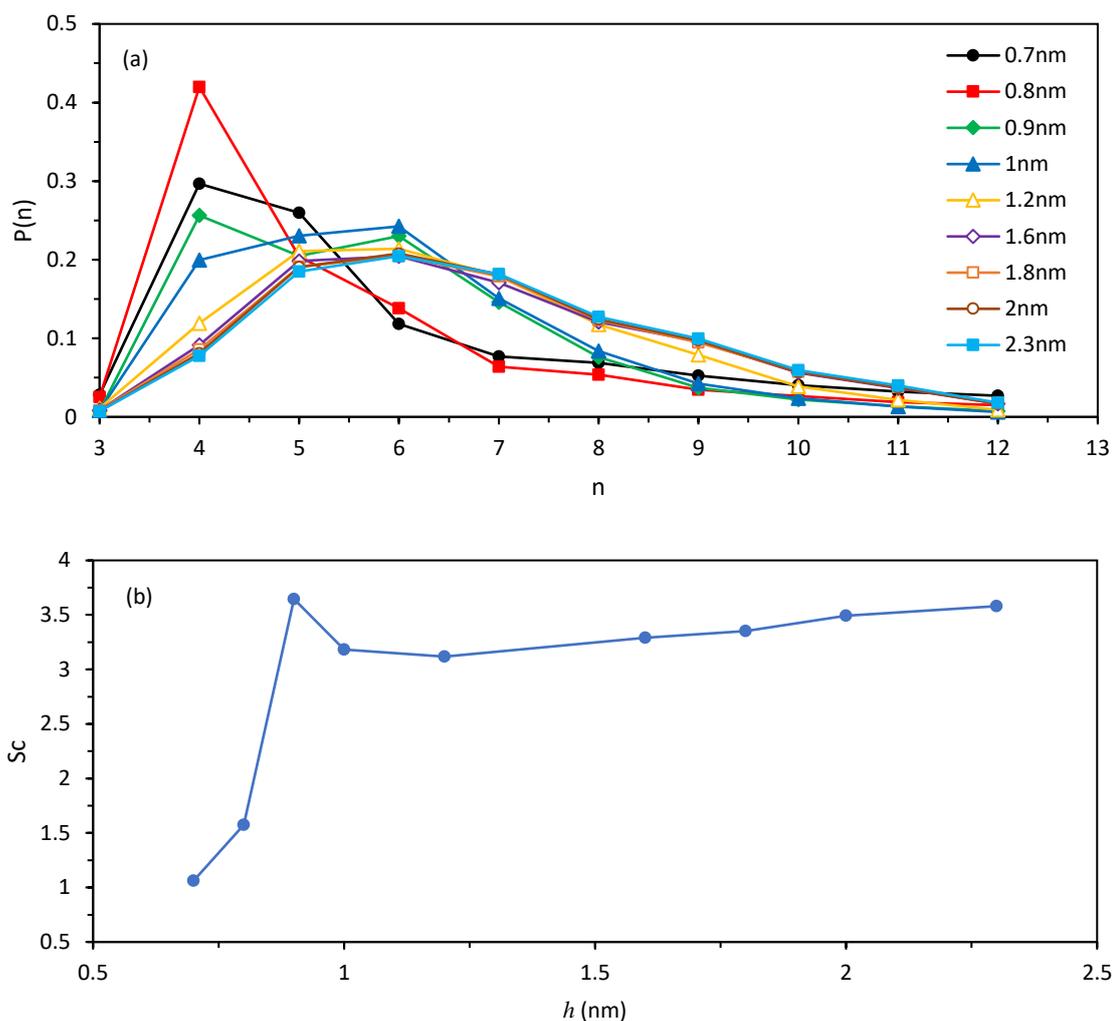

*Figure 8 (a). The distribution of rings (n) as a function of the degree of confinement (h=0.7-2.3nm). (b) The configurational entropy Sc as a function of the confinement.*

## 4. Conclusions

In summary, we conducted molecular dynamics simulations on a number of rigid graphite slits characterised by different heights ($h$) to investigate whether the graphite wettability changes as a function of the degree of confinement. In our simulations, we do not impose any density within the slits, which are filled by the water of the reservoirs they are in contact with. To assess the change in wettability and its impact on the flow properties, we calculated both the solid/liquid interfacial tension and friction coefficients from which we obtained the slip lengths. Our calculations show that large oscillations in the value of all these interfacial properties occur at strong and intermediate confinements and peak at $h$=0.9nm when the surface becomes super-hydrophobic. For larger confinements, all the properties converge to constant values. The friction coefficients (and slip lengths) plateau to the bulk value for interlayer distance larger than 2.0nm, however the surface tension values oscillate until larger confinement levels and converge only for interlayer distance larger than 2.6nm. We found that, under strong confinement, the value of the interfacial tension is correlated to that of the interfacial density and that, for all confinements, the number of water layers formed in the channel affects its value. The surface acquires its highest surface tension when the water molecules in the slit form two distinct layers (channel height 0.9nm), and again when a third layer (channel height 1.2nm) and a fifth layer (channel height 2nm) appear. For these three channel heights the averaged normal pressure within the slit is at its highest. The friction coefficient (and slip length) also reaches its highest (lowest) value at these confinement levels but $\gamma$ (and $\lambda$) reaches the bulk value for $h$=2.0nm. During the structural analysis, we concluded that the water structure is different under strong confinement compared to the larger channel heights. In particular, the structural analysis of the hydrogen bond network of the interfacial layer clarifies what happens at the interlayer distance of 0.9nm, which seems to be a key geometrical parameter at which interfacial tension, friction and slip length show the most remarkable difference compared to bulk. The analysis shows that at this degree of confinement a transition (of first order) from a 2D to a 3D network geometry takes place and the entropy of the hydrogen bond network suddenly increases, the lateral diffusivity decreases, and the water molecules start to flip from one layer to the other.

This oscillatory behaviour in interfacial tension, the friction, and the slip length under confinement can be used to develop more reliable macroscopic models to predict water flux in nanochannel and shed light on the wettability of the solid surfaces under nanoconfinement.

**Acknowledgements**

The authors would also like to acknowledge the assistance given by IT Services and the use of the Computational Shared Facility at the University of Manchester. The authors would also like to thank Dr Chris D. Williams, Dr. Gerardo Villabona-campos and Dr. Carlos Avendaño for useful discussions. MC, JDE and PC thank the European Union's Horizon 2020 research and innovation programme project VIMMP under Grant Agreement No. 760907.

**Author Contributions:**

**ORCID**

Zixuan Wei: 0000-0002-8854-279X

Mara Chiricotto: 0000-0003-1609-5254

Joshua D. Elliott: 0000-0002-0729-246X

Fausto Martelli: 0000-0002-5350-8225

Paola Carbone: 0000-0001-9927-8376

**Competing Interests**

The authors have no competing financial interests.

**Materials and Correspondence**

Supporting information including figures.